\documentclass[a4paper,11pt]{article}
\usepackage{pos}

\title{New results on $|V_{ub}|$ using inclusive and exclusive $B$ decays from the Belle experiment}

\author*[a]{Lu Cao}

\affiliation{Deutsches Elektronen-Synchrotron (DESY),\\
 Notkestraße 85, Hamburg, Germany}
  
\note{On behalf of the Belle collaborations.}

\emailAdd{lu.cao@desy.de}

\abstract{
We present two recent measurements of semileptonic B decays at Belle, offering valuable insights into the determination of the Cabibbo-Kobayashi-Maskawa (CKM) matrix element $|V_{ub}|$. These analyses use the full Belle dataset, comprising $772 \times 10^6$ $B\bar{B}$ pairs collected at the $\Upsilon(4S)$ resonance. With an innovative strategy, the inclusive $B \to X_u  \ell \nu$ and exclusive $B \to \pi  \ell \nu$ decays are simultaneously analyzed for the first time, and the $|V_{ub}|$ ratio is extracted as $|V_{ub}^{\mathrm{excl.}}|/|V_{ub}^{\mathrm {incl.}}|=0.97 \pm 0.12$. Furthermore, we provide preliminary results for the inclusive branching fractions ratio of $B \to X_u  \ell \nu$ and $B \to X_c  \ell \nu$ decays, accompanied by additional interpretations aimed at deriving the inclusive $|V_{ub}|/|V_{cb}|$ ratio.}

\FullConference{The European Physical Society Conference on High Energy Physics (EPS-HEP2023)\\
 21-25 August 2023\\
Hamburg, Germany\\}

\begin{document}
\maketitle

\section{Introduction}
In the Standard Model of particle physics (SM), the CKM matrix \cite{Cabibbo:1963yz, Kobayashi:1973fv} describes the quark mixing and accounts for $CP-$violation in the quark sector. One of the crucial tests of the SM is the precise determination of the magnitude of the matrix elements. In $b-$flavor scope, the corresponding world averages of $|V_{xb}|$ from exclusive and inclusive determinations \cite{HFLAV:2022pwe} are 

\begin{equation}
\label{eq:avr-vub}
\left|V_{u b}^{\text {excl. }}\right|=(3.51  \pm 0.12) \times 10^{-3} , \; \left|V_{u b}^{\text {incl. }}\right|=\left(4.19 \pm 0.16 \right) \times 10^{-3} ,  \; \left|V_{c b}^{\text {incl. }}\right|=\left(42.19 \pm 0.78 \right) \times 10^{-3} , 
\end{equation}

\noindent and their ratios are

\begin{equation}
\label{eq:avr-vubratio}
\begin{array}{l}

\left|V_{ub}^{\mathrm{excl.}}\right|/\left|V_{ub}^{\mathrm {incl.}}\right|=0.84 \pm 0.04, \\
\left|V_{ub}^{\mathrm{incl.}}\right|/\left|V_{cb}^{\mathrm {incl.}}\right|=0.099 \pm 0.004.
\end{array}
\end{equation}

\noindent The first ratio reflecting the disagreement between inclusive and exclusive $|V_{u b}|$ is about 3.7 standard deviations from the unity. The second ratio involves only the inclusive decays and combines the $b \to u \ell \nu$ and $b \to c \ell \nu$ decays. Experimentally, many systematic uncertainties such as the tagging calibration or the lepton identification uncertainties can be cancelled in such ratio measurements and one can directly test the SM expectation.

\section{Ratio of exclusive and inclusive $|V_{ub}|$}

The first simultaneous determination of $|V_{ub}|$ using inclusive and exclusive decays is performed at the Belle experiment~\cite{jointVub}. The hadronic decays of one of the $B$ mesons are reconstructed via the full reconstruction algorithm \cite{Feindt:2011mr} based on neural networks. The event reconstruction strategies are inherited from the recent Belle measurements ~\cite{cao:2021prd, cao:2021prl}, where all tracks and clusters not used in the construction of the $B_{\mathrm{tag}}$ candidate are used to reconstruct the signal side. The four-momentum of hadronic system $p_{X}$ is defined as a sum of the four-momenta of tracks and clusters which are not involved in reconstructing the $B_{\mathrm{tag}}$ and signal lepton. With the fully reconstructed four-momentum of $B_{\mathrm{tag}}$ and the known beam-momentum, the four-momentum of signal $B$ can be defined. 

To separate the signal $B\to X_{u} \ell \nu$ decay from the $B\to X_{c} \ell \nu$ events, a machine learning based classification with boosted decision trees (BDTs) is utilized. Furthermore, we apply a selection on the thrust  $T = \mathrm{max}_{|\mathbf{n}| = 1} \left( \sum_i | \mathbf{p_i} \cdot \mathbf{n}| / \sum_i  | \mathbf{p_i} |  \right)$ to enhance the $B \to \pi  \ell \nu$ purity. After all selections, a two-dimensional fit is employed to disentangle exclusive $B \to \pi  \ell \nu$ decays from other inclusive $B \to X_u  \ell \nu$ events and backgrounds. This fitting approach takes into account the number of charged pions $N_{\pi^\pm}$ in the hadronic $X_u$ system and the four-momentum transfer $q^2$ between the $B$ and $X_u$ system. We constrain the Bourrely-Caprini-Lellouch (BCL) expansion coefficients of $B \to \pi \ell \nu$ form factors to the LQCD values of Ref.~\cite{FLAG:2021npn}, combining LQCD calculations from several groups~\cite{FermilabLattice:2015mwy,Flynn:2015mha}. The additive and multiplicative systematic uncertainties are considered in the likelihood fit by adding bin-wise nuisance parameters for each template. The parameters are constrained to a multivariate Gaussian distribution with a covariance reflecting the sum of all considered systematic effects, and the correlation structure between templates originating from shared sources is taken into account. Various fit scenarios are applied to check the consistency of the nominal results. We fit with combined or separate normalizations of the $B \to \pi^{-}  \ell^{+} \nu$ and $B \to \pi^{0}  \ell^{+} \nu$ decays and also test with including only LQCD or LQD combined with external experimental constraints on the BCL form factors ~\cite{FLAG:2021npn}. Fig.~\ref{fig:sig} shows the $(q^2:N_{\pi^\pm})$ distribution of the signal region after the fit and with only using LQCD information.

\begin{figure}[h!]
	\centering
		 \includegraphics[width=0.58\linewidth]{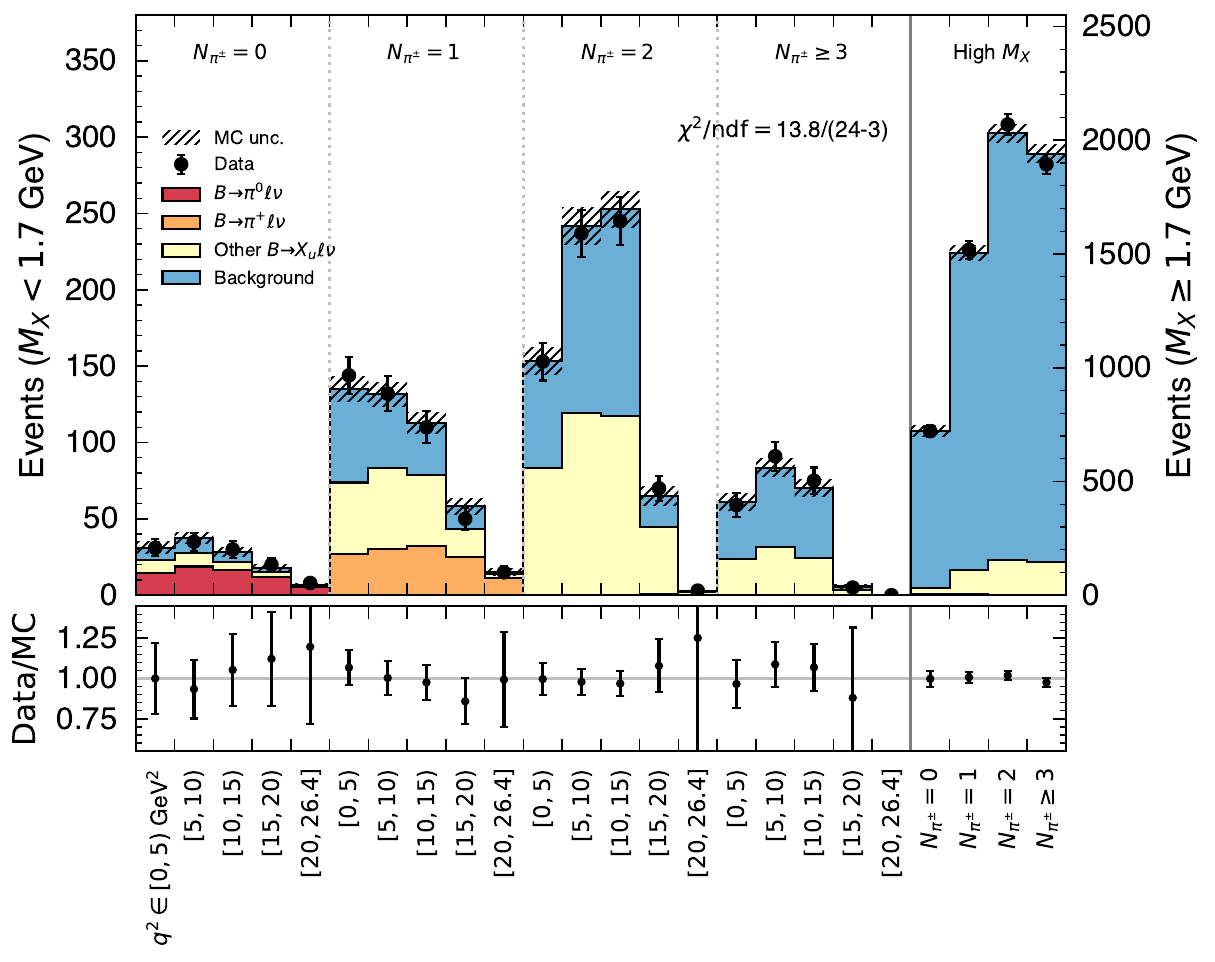} 
	\caption{\label{fig:sig} 
	The $q^2:N_{\pi^\pm}$ spectrum after the 2D fit is shown for the scenario that only uses LQCD information. 
	} 
\end{figure}

\begin{figure}[h!]
	\centering
		 \includegraphics[width=0.48\linewidth]{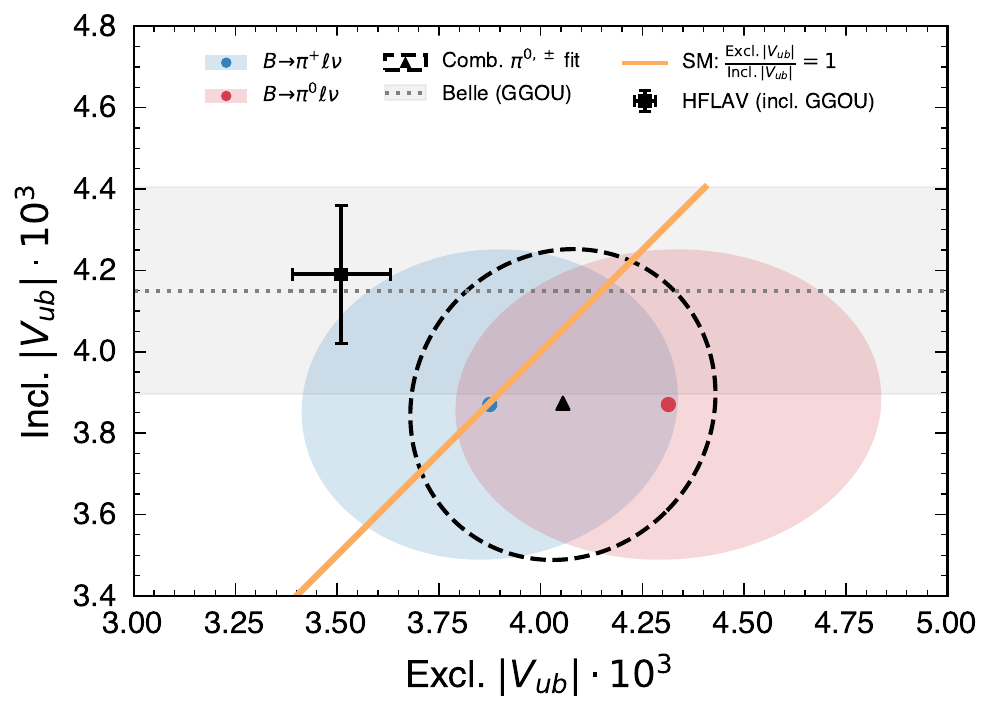}  
		 \includegraphics[width=0.48\linewidth]{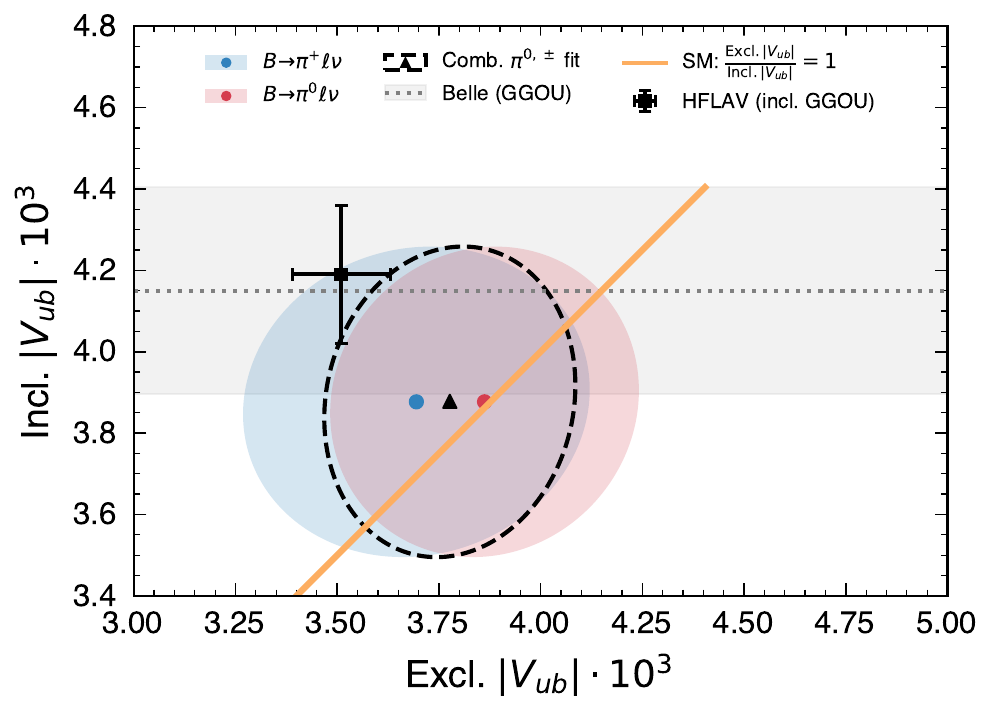} 		 
	
	\caption{ 
	The $|V_{ub}|$\ values obtained with the fits using (top) LQCD or (bottom) LQCD and experimental constraints for the $\overline B^0 \to \pi^+ \ell^- \bar \nu_\ell$ form factor are shown. The inclusive $|V_{ub}|$ value is based on the decay rate from the GGOU calculation~\cite{GGOU}. The values obtained from the previous Belle measurement ~\cite{cao:2021prd} (grey band) and the world averages from Ref.~\cite{HFLAV:2022pwe} (black marker) are also shown. The shown ellipses correspond to 39.3\% confidence levels ($\Delta \chi^2 = 1$).
	}
\label{fig:npi-vub-FLAG} 	
\end{figure}

The measured $B \to \pi^{-}  \ell^{+} \nu$ and $B \to \pi^{0}  \ell^{+} \nu$ yields are corrected for efficiency effects to determine the corresponding branching fractions. The measured inclusive yield is calculated from the sum of $B \to \pi^{-}  \ell^{+} \nu$, $B \to \pi^{0}  \ell^{+} \nu$, and other $B \to X_{u}  \ell \nu$ events and unfolded to correspond to a partial branching fraction $\Delta\mathcal{B}$ with $E_\ell^B > 1.0 \, \mathrm{GeV}$. Using calculations for the inclusive partial rate and the fitted form factor parameters, we can determine values for $|V_{ub}|$. The results of $|V_{ub}^{\mathrm{excl.}}|/|V_{ub}^{\mathrm {incl.}}|$ obtained in various fit scenarios are illustrated in Fig.~\ref{fig:npi-vub-FLAG}.

Our findings indicate ratios that align closely with this expectation but exhibit a deviation of 1.5 standard deviations from the ratio of the current world averages of inclusive and exclusive $|V_{ub}|$. This tension decreases to 1.2 standard deviations when including the constraint based on the full theoretical and experimental knowledge of the $B \to \pi  \ell \nu$ form factor shape. With this setup, we obtain the results on $\left|V_{ub}^{\mathrm{excl.}} \right| = (3.78 \pm 0.23 \pm 0.16 \pm 0.14)\times 10^{-3}$ and $\left|V_{ub}^{\mathrm{incl.}} \right| = (3.88 \pm 0.20 \pm 0.31 \pm 0.09)\times 10^{-3}$, with the uncertainties being the statistical, systematic, and theoretical errors. The ratio of $\left|V_{ub}^{\mathrm{excl.}} \right| / \left|V_{ub}^{\mathrm{incl.}} \right| = 0.97 \pm 0.12$ is compatible with unity. Moreover, the averaged $|V_{ub}|$ derived from our inclusive and exclusive values, using LQCD and additional experimental information, is $|V_{ub}|  =  (3.84 \pm 0.26)\times 10^{-3}$. This result is compatible with the expectation from CKM unitarity of $|V_{ub}^{\mathrm{CKM}}| = (3.64 \pm 0.07)\times 10^{-3}$~\cite{CKMfitter2021} within 0.8 standard deviations.

\section{Ratio of partial branching fractions of $B \to X_u  \ell \nu$ and $B \to X_c  \ell \nu$}
The semileptonic inclusive decays $B \to X_u  \ell \nu$ and $B \to X_c  \ell  \nu$ are analyzed with the Belle~II hadronic tagging algorithm \cite{Keck:2018lcd}. The event with sizeable missing four-momentum and partially-reconstructed $D^{*}$ candidates are rejected. In addition, the signal lepton with $E_{\ell}^{B}=\left|\mathbf{p}_{\ell}^{\mathbf{B}}\right|>1$ GeV in the signal-$B$ rest frame is used to identify the semileptonic decays.

\begin{figure}[h!]
	\centering
		 \includegraphics[width=0.85\linewidth]{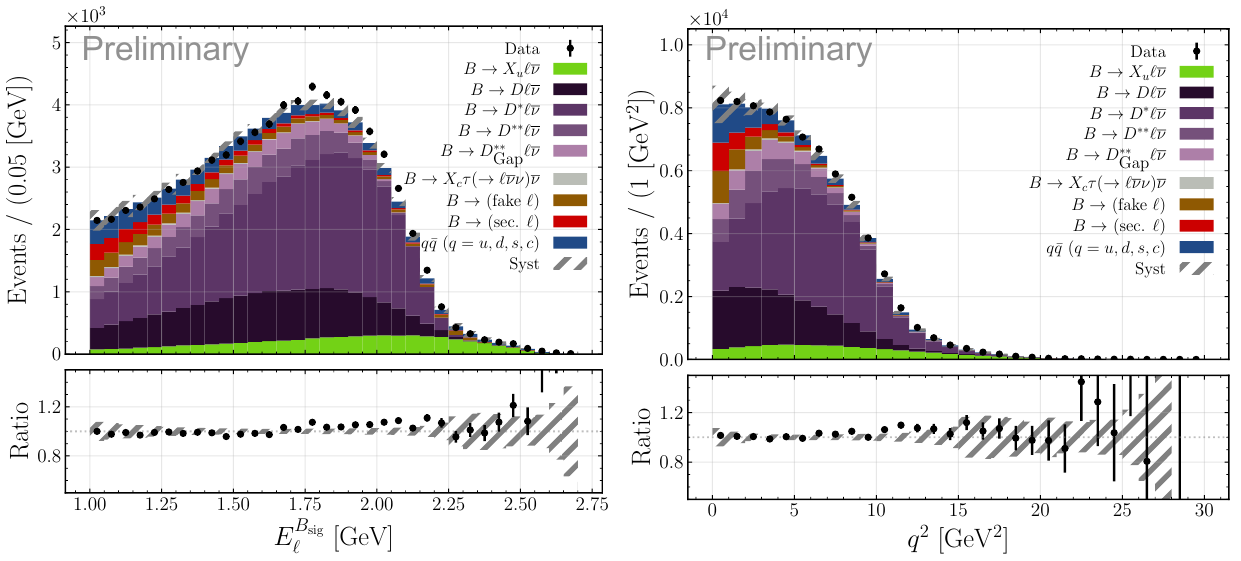} 
          \includegraphics[width=0.85\linewidth]{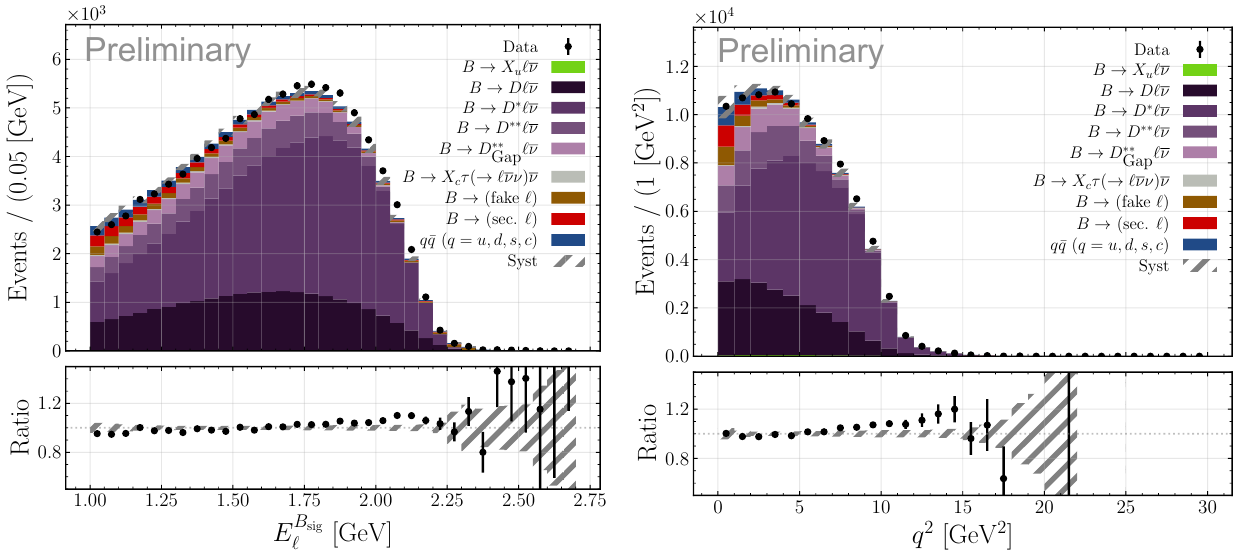} 
	\caption{\label{fig:enhance} 
	The $E_{\ell}^{B_{\mathrm{sig}}}$ and $q^{2}$ spectra for the $B \to X_u  \ell \nu$ enhanced (top) and depleted (bottom) sub-samples. 
	} 
\end{figure}

After the selections, the $B \to X_u  \ell \nu$ enhanced sub-sample is categorized by observing an even value of kaons in the hadronic system $(N_{K^{\pm}}+N_{K_{S}^{0}})$. The remaining events are defined as $B \to X_u  \ell \nu$ depleted sub-sample, which is dominant by $B \to X_c  \ell \nu$ decays and used to modify the modeling of $B \to X_c  \ell \nu$ in a data-driven way to minimize potential bias due to limited knowledge of this component. Fig.~\ref{fig:enhance} shows the distributions of the reconstructed lepton energy in the B meson rest frame $E_{\ell}^{B}$ and the squared four-momentum transfer $q^{2}$ in the $B \to X_u  \ell \nu$ enhanced and depleted sub-samples, respectively. To derive the data-driven template of $B \to X_c  \ell \nu$, the simulated continuum contribution is calibrated to the measured off-resonances dataset and the background events with secondary or fake leptons are also calibrated in the control regions. 

The $B \to X_u  \ell  \nu$ yields are extracted in a two-dimensional fit on $q^{2}$ and $E_{\ell}^{B}$. The fit result is shown in Fig.~\ref{fig:sig}. The $B \to X_c  \ell  \nu$ yields are obtained by subtracting other contributions in the total $B \to X \ell \nu$ sample. The measured yields are unfolded using the Singular Value Decomposition (SVD) algorithm of Ref.~\cite{svd}.

\begin{figure}[h!]
	\centering
		 \includegraphics[width=0.75\linewidth]{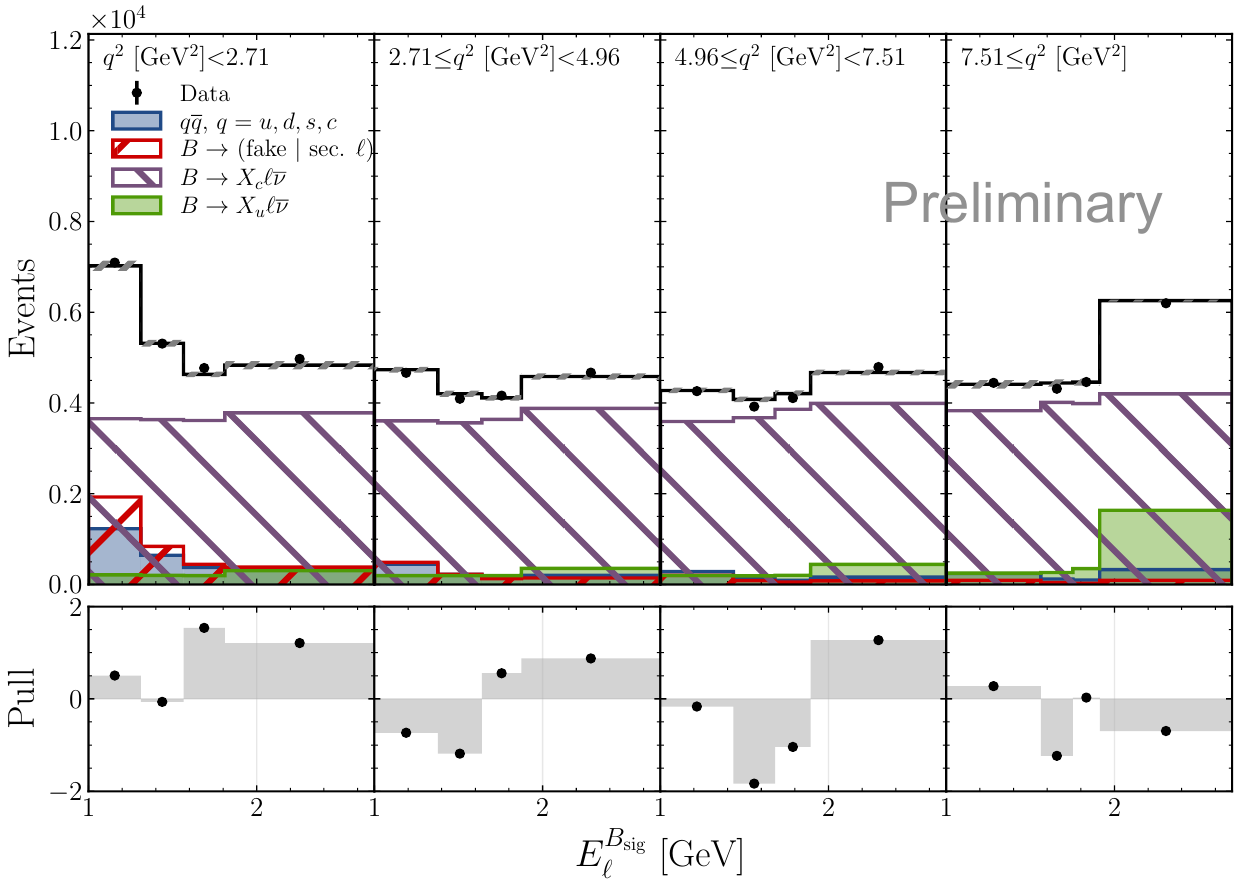} 
	\caption{\label{fig:sig} 
	The $q^{2}: E_{\ell}^{B_{\mathrm{sig}}}$ spectrum after the $B \to X_u  \ell \nu$ fit. 
	} 
\end{figure}

The study focused on the partial phase space region where $E_{\ell}^{B} > 1 \, \mathrm{GeV}$, encompassing fractions of $\epsilon^{u}_{\Delta} = 86\%$ and $\epsilon^{c}_{\Delta} = 79\%$ of the $B \to X_u  \ell \nu$ and $B \to X_c  \ell \nu$ decays, respectively. The preliminary result for the partial branching faction ratio is $\Delta\mathcal{B}(B \to X_{u}\ell\nu)/\Delta\mathcal{B}(B \to X_{c}\ell\nu) = 1.95(1 \pm 8.4\%_{\mathrm{stat}} \pm 7.8\%_{\mathrm{syst}})\times 10^{-2}$. This ratio provides insight into the inclusive $|V_{ub}|/|V_{cb}|$ ratio, with theoretical inputs of partial decay rates for both decays. Furthermore, by taking the external normalization of $\Delta \mathcal{B}(B \to X_c \ell  \nu)$, we find the resulting $|V_{ub}|$ value agrees very well with the previous Belle inclusive measurement~\cite{cao:2021prd}.

\section{Summary}

In summary, we report the two new measurements from the Belle experiment to examine the long-standing discrepancy between the inclusive and exclusive determinations of $|V_{ub}|$. The ratios of $|V_{ub}|^{\mathrm{excl}}/|V_{ub}|^{\mathrm{incl}}$ and $|V_{ub}|^{\mathrm{incl}}/|V_{cb}|^{\mathrm{incl}}$ are measured for the first time. While these findings offer valuable insights, the puzzle remains unsolved due to the current limitations in experimental precision. The anticipated large data set of Belle~II will be beneficial for exploring this topic further through the presented innovative approaches.

\bibliographystyle{apsrev4-1}
\bibliography{ref}

\begin{thebibliography}{14}%
\makeatletter
\providecommand \@ifxundefined [1]{%
 \@ifx{#1\undefined}
}%
\providecommand \@ifnum [1]{%
 \ifnum #1\expandafter \@firstoftwo
 \else \expandafter \@secondoftwo
 \fi
}%
\providecommand \@ifx [1]{%
 \ifx #1\expandafter \@firstoftwo
 \else \expandafter \@secondoftwo
 \fi
}%
\providecommand \natexlab [1]{#1}%
\providecommand \enquote  [1]{``#1''}%
\providecommand \bibnamefont  [1]{#1}%
\providecommand \bibfnamefont [1]{#1}%
\providecommand \citenamefont [1]{#1}%
\providecommand \href@noop [0]{\@secondoftwo}%
\providecommand \href [0]{\begingroup \@sanitize@url \@href}%
\providecommand \@href[1]{\@@startlink{#1}\@@href}%
\providecommand \@@href[1]{\endgroup#1\@@endlink}%
\providecommand \@sanitize@url [0]{\catcode `\\12\catcode `\$12\catcode
  `\&12\catcode `\#12\catcode `\^12\catcode `\_12\catcode `\%12\relax}%
\providecommand \@@startlink[1]{}%
\providecommand \@@endlink[0]{}%
\providecommand \url  [0]{\begingroup\@sanitize@url \@url }%
\providecommand \@url [1]{\endgroup\@href {#1}{\urlprefix }}%
\providecommand \urlprefix  [0]{URL }%
\providecommand \Eprint [0]{\href }%
\providecommand \doibase [0]{http://dx.doi.org/}%
\providecommand \selectlanguage [0]{\@gobble}%
\providecommand \bibinfo  [0]{\@secondoftwo}%
\providecommand \bibfield  [0]{\@secondoftwo}%
\providecommand \translation [1]{[#1]}%
\providecommand \BibitemOpen [0]{}%
\providecommand \bibitemStop [0]{}%
\providecommand \bibitemNoStop [0]{.\EOS\space}%
\providecommand \EOS [0]{\spacefactor3000\relax}%
\providecommand \BibitemShut  [1]{\csname bibitem#1\endcsname}%
\let\auto@bib@innerbib\@empty
\bibitem [{\citenamefont {Cabibbo}(1963)}]{Cabibbo:1963yz}%
  \BibitemOpen
  \bibfield  {author} {\bibinfo {author} {\bibfnamefont {N.}~\bibnamefont
  {Cabibbo}},\ }\href {\doibase 10.1103/PhysRevLett.10.531} {\bibfield
  {journal} {\bibinfo  {journal} {Phys. Rev. Lett.}\ }\textbf {\bibinfo
  {volume} {10}},\ \bibinfo {pages} {531} (\bibinfo {year} {1963})}\BibitemShut
  {NoStop}%
\bibitem [{\citenamefont {Kobayashi}\ and\ \citenamefont
  {Maskawa}(1973)}]{Kobayashi:1973fv}%
  \BibitemOpen
  \bibfield  {author} {\bibinfo {author} {\bibfnamefont {M.}~\bibnamefont
  {Kobayashi}}\ and\ \bibinfo {author} {\bibfnamefont {T.}~\bibnamefont
  {Maskawa}},\ }\href {\doibase 10.1143/PTP.49.652} {\bibfield  {journal}
  {\bibinfo  {journal} {Prog. Theor. Phys.}\ }\textbf {\bibinfo {volume}
  {49}},\ \bibinfo {pages} {652} (\bibinfo {year} {1973})}\BibitemShut
  {NoStop}%
\bibitem [{\citenamefont {Amhis}\ \emph {et~al.}(2023)\citenamefont {Amhis}
  \emph {et~al.}}]{HFLAV:2022pwe}%
  \BibitemOpen
  \bibfield  {author} {\bibinfo {author} {\bibfnamefont {Y.~S.}\ \bibnamefont
  {Amhis}} \emph {et~al.} (\bibinfo {collaboration} {HFLAV}),\ }\href {\doibase
  10.1103/PhysRevD.107.052008} {\bibfield  {journal} {\bibinfo  {journal}
  {Phys. Rev. D}\ }\textbf {\bibinfo {volume} {107}},\ \bibinfo {pages}
  {052008} (\bibinfo {year} {2023})},\ \Eprint
  {http://arxiv.org/abs/2206.07501} {arXiv:2206.07501 [hep-ex]} \BibitemShut
  {NoStop}%
\bibitem [{\citenamefont {Cao}\ \emph {et~al.}(2023)\citenamefont {Cao} \emph
  {et~al.}}]{jointVub}%
  \BibitemOpen
  \bibfield  {author} {\bibinfo {author} {\bibfnamefont {L.}~\bibnamefont
  {Cao}} \emph {et~al.} (\bibinfo {collaboration} {Belle Collaboration}),\
  }\href@noop {} {\  (\bibinfo {year} {2023})},\ \bibinfo {note} {accepted by
  Phys. Rev. Lett.},\ \Eprint {http://arxiv.org/abs/2303.17309}
  {arXiv:2303.17309 [hep-ex]} \BibitemShut {NoStop}%
\bibitem [{\citenamefont {Feindt}\ \emph {et~al.}(2011)\citenamefont {Feindt},
  \citenamefont {Keller}, \citenamefont {Kreps}, \citenamefont {Kuhr},
  \citenamefont {Neubauer}, \citenamefont {Zander},\ and\ \citenamefont
  {Zupanc}}]{Feindt:2011mr}%
  \BibitemOpen
  \bibfield  {author} {\bibinfo {author} {\bibfnamefont {M.}~\bibnamefont
  {Feindt}}, \bibinfo {author} {\bibfnamefont {F.}~\bibnamefont {Keller}},
  \bibinfo {author} {\bibfnamefont {M.}~\bibnamefont {Kreps}}, \bibinfo
  {author} {\bibfnamefont {T.}~\bibnamefont {Kuhr}}, \bibinfo {author}
  {\bibfnamefont {S.}~\bibnamefont {Neubauer}}, \bibinfo {author}
  {\bibfnamefont {D.}~\bibnamefont {Zander}}, \ and\ \bibinfo {author}
  {\bibfnamefont {A.}~\bibnamefont {Zupanc}},\ }\href {\doibase
  10.1016/j.nima.2011.06.008} {\bibfield  {journal} {\bibinfo  {journal} {Nucl.
  Instrum. Meth. A}\ }\textbf {\bibinfo {volume} {654}},\ \bibinfo {pages}
  {432} (\bibinfo {year} {2011})}\BibitemShut {NoStop}%
\bibitem [{\citenamefont {Cao}\ \emph {et~al.}(2021{\natexlab{a}})\citenamefont
  {Cao} \emph {et~al.}}]{cao:2021prd}%
  \BibitemOpen
  \bibfield  {author} {\bibinfo {author} {\bibfnamefont {L.}~\bibnamefont
  {Cao}} \emph {et~al.} (\bibinfo {collaboration} {Belle Collaboration}),\
  }\href {\doibase 10.1103/PhysRevD.104.012008} {\bibfield  {journal} {\bibinfo
   {journal} {Phys. Rev. D}\ }\textbf {\bibinfo {volume} {104}},\ \bibinfo
  {pages} {012008} (\bibinfo {year} {2021}{\natexlab{a}})},\ \Eprint
  {http://arxiv.org/abs/2102.00020} {arXiv:2102.00020 [hep-ex]} \BibitemShut
  {NoStop}%
\bibitem [{\citenamefont {Cao}\ \emph {et~al.}(2021{\natexlab{b}})\citenamefont
  {Cao} \emph {et~al.}}]{cao:2021prl}%
  \BibitemOpen
  \bibfield  {author} {\bibinfo {author} {\bibfnamefont {L.}~\bibnamefont
  {Cao}} \emph {et~al.} (\bibinfo {collaboration} {Belle Collaboration}),\
  }\href {\doibase 10.1103/PhysRevLett.127.261801} {\bibfield  {journal}
  {\bibinfo  {journal} {Phys. Rev. Lett.}\ }\textbf {\bibinfo {volume} {127}},\
  \bibinfo {pages} {261801} (\bibinfo {year} {2021}{\natexlab{b}})},\ \Eprint
  {http://arxiv.org/abs/2107.13855} {arXiv:2107.13855 [hep-ex]} \BibitemShut
  {NoStop}%
\bibitem [{\citenamefont {Aoki}\ \emph {et~al.}(2022)\citenamefont {Aoki} \emph
  {et~al.}}]{FLAG:2021npn}%
  \BibitemOpen
  \bibfield  {author} {\bibinfo {author} {\bibfnamefont {Y.}~\bibnamefont
  {Aoki}} \emph {et~al.} (\bibinfo {collaboration} {Flavour Lattice Averaging
  Group (FLAG)}),\ }\href {\doibase 10.1140/epjc/s10052-022-10536-1} {\bibfield
   {journal} {\bibinfo  {journal} {Eur. Phys. J. C}\ }\textbf {\bibinfo
  {volume} {82}},\ \bibinfo {pages} {869} (\bibinfo {year} {2022})},\ \Eprint
  {http://arxiv.org/abs/2111.09849} {arXiv:2111.09849 [hep-lat]} \BibitemShut
  {NoStop}%
\bibitem [{\citenamefont {Bailey}\ \emph {et~al.}(2015)\citenamefont {Bailey}
  \emph {et~al.}}]{FermilabLattice:2015mwy}%
  \BibitemOpen
  \bibfield  {author} {\bibinfo {author} {\bibfnamefont {J.~A.}\ \bibnamefont
  {Bailey}} \emph {et~al.} (\bibinfo {collaboration} {Fermilab Lattice and MILC
  Collaborations}),\ }\href {\doibase 10.1103/PhysRevD.92.014024} {\bibfield
  {journal} {\bibinfo  {journal} {Phys. Rev. D}\ }\textbf {\bibinfo {volume}
  {92}},\ \bibinfo {pages} {014024} (\bibinfo {year} {2015})},\ \Eprint
  {http://arxiv.org/abs/1503.07839} {arXiv:1503.07839 [hep-lat]} \BibitemShut
  {NoStop}%
\bibitem [{\citenamefont {Flynn}\ \emph {et~al.}(2015)\citenamefont {Flynn},
  \citenamefont {Izubuchi}, \citenamefont {Kawanai}, \citenamefont {Lehner},
  \citenamefont {Soni}, \citenamefont {Van~de Water},\ and\ \citenamefont
  {Witzel}}]{Flynn:2015mha}%
  \BibitemOpen
  \bibfield  {author} {\bibinfo {author} {\bibfnamefont {J.~M.}\ \bibnamefont
  {Flynn}}, \bibinfo {author} {\bibfnamefont {T.}~\bibnamefont {Izubuchi}},
  \bibinfo {author} {\bibfnamefont {T.}~\bibnamefont {Kawanai}}, \bibinfo
  {author} {\bibfnamefont {C.}~\bibnamefont {Lehner}}, \bibinfo {author}
  {\bibfnamefont {A.}~\bibnamefont {Soni}}, \bibinfo {author} {\bibfnamefont
  {R.~S.}\ \bibnamefont {Van~de Water}}, \ and\ \bibinfo {author}
  {\bibfnamefont {O.}~\bibnamefont {Witzel}},\ }\href {\doibase
  10.1103/PhysRevD.91.074510} {\bibfield  {journal} {\bibinfo  {journal} {Phys.
  Rev. D}\ }\textbf {\bibinfo {volume} {91}},\ \bibinfo {pages} {074510}
  (\bibinfo {year} {2015})},\ \Eprint {http://arxiv.org/abs/1501.05373}
  {arXiv:1501.05373 [hep-lat]} \BibitemShut {NoStop}%
\bibitem [{\citenamefont {Gambino}\ \emph {et~al.}(2007)\citenamefont
  {Gambino}, \citenamefont {Giordano}, \citenamefont {Ossola},\ and\
  \citenamefont {Uraltsev}}]{GGOU}%
  \BibitemOpen
  \bibfield  {author} {\bibinfo {author} {\bibfnamefont {P.}~\bibnamefont
  {Gambino}}, \bibinfo {author} {\bibfnamefont {P.}~\bibnamefont {Giordano}},
  \bibinfo {author} {\bibfnamefont {G.}~\bibnamefont {Ossola}}, \ and\ \bibinfo
  {author} {\bibfnamefont {N.}~\bibnamefont {Uraltsev}},\ }\href {\doibase
  10.1088/1126-6708/2007/10/058} {\bibfield  {journal} {\bibinfo  {journal}
  {JHEP}\ }\textbf {\bibinfo {volume} {10}},\ \bibinfo {pages} {058} (\bibinfo
  {year} {2007})}\BibitemShut {NoStop}%
\bibitem [{\citenamefont {Charles}\ \emph {et~al.}(2005)\citenamefont {Charles}
  \emph {et~al.}}]{CKMfitter2021}%
  \BibitemOpen
  \bibfield  {author} {\bibinfo {author} {\bibfnamefont {J.}~\bibnamefont
  {Charles}} \emph {et~al.} (\bibinfo {collaboration} {CKMfitter Group}),\
  }\href {\doibase 10.1140/epjc/s2005-02169-1} {\bibfield  {journal} {\bibinfo
  {journal} {Eur. Phys. J. C}\ }\textbf {\bibinfo {volume} {41}},\ \bibinfo
  {pages} {1} (\bibinfo {year} {2005})},\ \bibinfo {note} {and updates of
  Spring 2021 on http://ckmfitter.in2p3.fr/},\ \Eprint
  {http://arxiv.org/abs/0406184} {arXiv:0406184 [hep-ph]} \BibitemShut
  {NoStop}%
\bibitem [{\citenamefont {Keck}\ \emph {et~al.}(2019)\citenamefont {Keck} \emph
  {et~al.}}]{Keck:2018lcd}%
  \BibitemOpen
  \bibfield  {author} {\bibinfo {author} {\bibfnamefont {T.}~\bibnamefont
  {Keck}} \emph {et~al.},\ }\href {\doibase 10.1007/s41781-019-0021-8}
  {\bibfield  {journal} {\bibinfo  {journal} {Comput. Softw. Big Sci.}\
  }\textbf {\bibinfo {volume} {3}},\ \bibinfo {pages} {6} (\bibinfo {year}
  {2019})},\ \Eprint {http://arxiv.org/abs/1807.08680} {arXiv:1807.08680
  [hep-ex]} \BibitemShut {NoStop}%
\bibitem [{\citenamefont {H{\"o}cker}\ and\ \citenamefont
  {Kartvelishvili}(1996)}]{svd}%
  \BibitemOpen
  \bibfield  {author} {\bibinfo {author} {\bibfnamefont {A.}~\bibnamefont
  {H{\"o}cker}}\ and\ \bibinfo {author} {\bibfnamefont {V.}~\bibnamefont
  {Kartvelishvili}},\ }\href {\doibase 10.1016/0168-9002(95)01478-0} {\bibfield
   {journal} {\bibinfo  {journal} {Nucl. Inst. Meth. A}\ }\textbf {\bibinfo
  {volume} {372}},\ \bibinfo {pages} {469} (\bibinfo {year}
  {1996})}\BibitemShut {NoStop}%
\end{thebibliography}%
\end{document}